\documentclass[12pt]{iopart}

\usepackage{amssymb}
\usepackage{graphicx}
\usepackage{mathrsfs,amsfonts}

\def\be{\begin{equation}}
\def\ee{\end{equation}}
\def\bea{\begin{eqnarray}}
\def\eea{\end{eqnarray}}

\begin{document}

\title[Nonlinear Resonance in Ho\v{r}ava-Lifshitz Bouncing Cosmologies]{Nonlinear Resonance in Ho\v{r}ava-Lifshitz Bouncing Cosmologies}

\author{Rodrigo Maier$^{1}$}

\address{Institute of Cosmology and Gravitation, University of Portsmouth,\\
Dennis Sciama Building, Portsmouth, PO1 3FX, United Kingdom}
\ead{$^{1}$rodrigo.maier@port.ac.uk}
\begin{abstract}
In this paper I examine the phase space dynamics in the framework of Non-Projectable Ho\v{r}ava-Lifshitz bouncing cosmologies.
By considering a closed Friedmann-Lema\^itre-Robertson-Walker
(FLRW) geometry, the first integral
contains a correction term that
leads to nonsingular metastable bounces in the early evolution of the universe. The
matter content of the model is a massive conformally coupled scalar field, dust and radiation.
A nonvanishing cosmological constant connected to a de Sitter attractor in the phase space
is also assumed.
In
narrow windows of the parameter space, labeled by an integer $n\geq 2$, nonlinear resonance
phenomena may destroy the KAM tori that trap the scalar field, leading to an exit
to the de Sitter attractor. As a consequence nonlinear resonance imposes constraints on the
parameters and in the initial configurations of the models so that an accelerated expansion may be
realized.
\end{abstract}


\maketitle

\section{Introduction}
Although General Relativity is the most successful theory that currently describes
gravitation, it presents some intrinsic crucial problems when one tries to construct a
cosmological model in accordance with observational
data. In cosmology, the $\Lambda {\rm CDM}$ model gives us important predictions concerning
the evolution of the universe and about its current state \cite{mukhanov}. However, let us
assume that the initial conditions of our Universe were fixed when the early
universe emerged from the semi-Planckian regime and started its classical
expansion. Evolving back such initial conditions using the Einstein field
equations, we see that our universe is driven towards an initial singularity
where the classical regime is no longer valid \cite{wald}.
\par
Notwithstanding the cosmic censorship conjecture \cite{penrose}, there is no doubt
that General Relativity must be properly corrected or even replaced by a
completely new theory, let us say a quantum theory of gravity. This demand
is in order to solve the issue of the presence of the initial singularity predicted by
classical General Relativity in the beginning of the
universe.
\par
One of the most important characteristics of our Universe supported by
observational data is its large scale of homogeneity and isotropy. However, when we consider
a homogeneous and isotropic model filled with baryonic matter, we find
several difficulties by taking into account the primordial state of our
Universe. Among such difficulties, we can mention the horizon and flatness
problems \cite{mukhanov}. Although the Inflationary
Paradigm\cite{abbott} allows one to solve problems like these, inflationary cosmology does not solve the problem of the initial
singularity.
\par
On the other hand,
since 1998 \cite{riess} observational data have been giving support to the
highly unexpected assumption that our Universe is currently in a state of accelerated expansion.
In order to explain this state of late-time acceleration, cosmologists have been considering the existence of some field
-- known as dark energy -- that violates the strong energy condition. Although it poses a problem to quantum
field theory on how to accommodate its observed value
with vacuum energy calculations\cite{ccp}, the cosmological constant seems to be the simplest and most
appealing candidate for dark energy. Therefore, nonsingular models which provide late-time acceleration should be strongly considered.
\par
During the last decades, bouncing models \cite{novello,ns} have been considered in
order to solve the problem of initial singularity predicted by General
Relativity. Such models (as \cite{nelson}) might provide attractive alternatives
to the inflationary paradigm
once they can solve the horizon and flatness problems, and justify the
power spectrum of primordial cosmological perturbations
inferred by observations.
\par
In 2009, P. Ho\v{r}ava proposed a
modified gravity theory by considering a Lifshitz-type anisotropic
scaling between space and time at high energies \cite{Horava}.
In this context, it has been shown \cite{cal, brand} that higher
spatial curvature terms can lead to regular bounce solutions in
the early universe. Since its proposal, several versions of Ho\v{r}ava-Lifshitz
gravity have emerged.


In the case of a $4$-dimensional ($1+3$) spacetime, the basic assumption which
is required by all the versions of Ho\v{r}ava-Lifshitz theories is
that a preferred foliation of spacetime
is {\it a priori} imposed. Therefore it is
natural to work with the Arnowitt-Deser-Misner (ADM) decomposition
of spacetime
\begin{eqnarray}
\label{eq0new}
ds^2=N^2dt^2-g_{ij}(N^i dt+dx^i)(N^j dt+dx^j),
\end{eqnarray}
where $N=N(t,x^i)$ is the lapse function, $N^i=N^i(t,x^i)$ is the shift and $g_{ij}=g_{ij}(t,x^i)$ is the spatial geometry. In this case the
final action of the theory will not be invariant under diffeomorphisms as in General Relativity. Nevertheless,
an invariant foliation preserving diffeomorphisms can be assumed. This is achieved if the action is invariant
under the symmetry of time reparametrization together with time-dependent spatial diffeomorphisms. That is:
\begin{eqnarray}
\label{eq1new}
t\rightarrow \bar{t}(t),~x^i\rightarrow\bar{x}^i(t,x^i).
\end{eqnarray}
It turns out that the only covariant object under spatial diffeomorphisms
that contains one time derivative of the spatial metric is the extrinsic curvature $K_{ij}$
\begin{eqnarray}
\label{eq2new}
K_{ij}=\frac{1}{2N}\Big[\frac{\partial g_{ij}}{\partial t}-\nabla_i N_j-\nabla_j N_i\Big]
\end{eqnarray}
where $\nabla_i$ is the covariant derivative built with the spatial metric $g_{ij}$.
Thus, to construct the general theory which is of second order in time derivatives, one needs to consider
the quadratic terms $K_{ij} K^{ij}$ and $K^2$ -- where $K$ is the trace of $K_{ij}$ -- in the extrinsic curvature.
By taking these terms into account we obtain the following general action
\begin{eqnarray}
\label{eq3new}
S\propto \int N \sqrt{-g}[K_{ij} K^{ij} - \lambda K^2 - U(g_{ij},N)] dx^3 dt
\end{eqnarray}
where $g$ is the
determinant of the spatial metric and $\lambda$ is a constant which corresponds to a dimensionless running coupling.
As in General Relativity the term $K_{ij}K^{ij}-K^2$ is invariant under four-dimensional diffeomorphisms,
we expect to recover the classical regime for $\lambda\rightarrow 1$. That is why it is a consensus that $\lambda$
must be a parameter sufficiently close to $1$.
In general, $U(g_{ij},N)$ can depend on the spatial metric and the lapse function because of the symmetry of the theory.
It is obvious that there are several invariant terms that one could include in $U$.
Particular choices resulted in different versions of Ho\v{r}ava-Lifshitz
gravity.
\par
Motivated by condensed matter systems, P. Ho\v{r}ava proposed a symmetry on $U$ that substantially reduces the
number of invariants\cite{Horava}. In this case, $U$ depends on a superpotential W given by the Chern-Simons term,
the curvature scalar and a term which mimics the cosmological constant.
It has been shown \cite{prob1} that this original assumption has to be broken if one intends to build a theory in
agreement to current
observations.
\par
The simplification $N=N(t)$ was also originally proposed by Ho\v{r}ava\cite{Horava}.
This condition defines a version of Ho\v{r}ava-Lifshitz gravity called Projectable. As $\partial N / \partial x^i \equiv 0$, the Projectable version also reduces the number of invariants that one can include in $U$.
The linearization of this version assuming a Minkowski background provides an extra scalar degree of freedom which
is classically unstable in the IR when $\lambda>1$ or $\lambda<1/3$, and is a ghost when $1/3<\lambda<1$ \cite{sotiriou2}. Although some physicists argue that
higher order derivatives can cut off these instabilities, it has been shown\cite{prob1,blas,bog,koyama} that a perturbative analysis is not consistent when $\lambda\rightarrow 1$ and the scalar mode gets strongly coupled. That is because the strongly coupled scale is unacceptably low. In this case, higher
order operators would modify the graviton dynamics at very low energies, being in conflict with current observations.
\par
Besides pure curvature invariants of $g_{ij}$, one may also include invariant contractions of $\partial (\ln{N})/\partial x^i$ in $U$. This assumption defines
the so-called Non-Projectable version of Ho\v{r}ava-Lifshitz gravity. Connected to the lowest order invariant $\partial_i (\ln{N})\partial^i (\ln{N})$, there is a parameter $\sigma$ which defines a ``safe'' domain of the theory\cite{sotiriou2,sotiriou1}. In fact, in this case there is also an extra scalar degree of freedom when one linearizes the theory
in a Minkowski background. However, when $0<\sigma<2$ and $\lambda>1$ this mode is not a ghost nor classically unstable (as long as detailed balance is not imposed).
Although the Non-Projectable version also exhibits a strong coupling\cite{prob1,sotiriou1,pap}, it has been argued that its scale is too high to be phenomenologically accessible from gravitational experiments\cite{sotiriou2}.
\par
In this paper I adhere to the so-called Non-Projectable Ho\v{r}ava-Lifshitz gravity in which I consider a nonsingular
FLRW cosmological model \cite{sotiriou1}.
The matter content is given by dust, radiation and a conformally coupled scalar field.
I also assume a nonvanishing cosmological constant in order to obtain a de Sitter atractor
in the phase space. In this context I show how an alternative exit to late-time acceleration (connected
to the de Sitter attractor) may be realized.
\par
In the next section I present a nonsingular homogeneous and isotropic cosmological model
-- sourced with perfect fluids, a cosmological constant, and a conformally coupled scalar field --
which arises from Non-Projectable Ho\v{r}ava-Lifshitz gravity. In section 3 I analyze the structure of the phase space.
In section 4 I restrict myself to the case of
dust and radiation in order to construct a simple model.
In section 5 I show how nonlinear resonance can provide an exit to the de Sitter attractor. In section 6
I exhibit the pattern of the resonance windows and show in which regions in the
parametric space late-time acceleration may be realized. Final remarks are given in the Conclusions.

\section{The Model}

Let us consider a model in which the matter content
is given by a nonminimally coupled massive scalar field $\phi$ and $N$ noninteracting perfect fluids
with equation of state $p_i=\omega_i \rho_i$ (i=1,.., N).
In this context, the $4$-D covariant Lagrangian ${\cal L}_m$ for the matter content can be written as
\begin{eqnarray}
\label{eq1}
{\cal L}_m = \sum^N_{i=1}{\cal L}_i-{\cal L}_{\phi},
\end{eqnarray}
where ${\cal L}_i$ are the Lagrangians for the noninteracting perfect fluids and
\begin{eqnarray}
\label{eq1.1}
{\cal L}_{\phi} = \frac{1}{2}\Big[(\phi_{,\alpha}\phi_{,\beta}g^{\alpha\beta}+m^2\phi^2)+\xi R\phi^2\Big],
\end{eqnarray}
with $R$ being the $4$-D Ricci scalar.
That is, ${\cal L}_m$ is the Lagrangian density of the massive
scalar field plus perfect fluids whose dynamics interact only with the metric $g_{\alpha\beta}$. We
further assume that the scalar field is nonminimally
coupled with $g_{\alpha\beta}$, with coupling parameter $\xi$.
\par
The fundamental symmetry assumed in Ho\v{r}ava--Lifshitz gravity (invariant under diffeomorphisms that preserve the foliation) provides enough gauge freedom to choose
\be
\label{choice}
N=1,\quad N^i=0,\quad g_{ij}=-a(t)^2\gamma_{ij}.
\ee
This puts the geometry (\ref{eq0new}) into the FLRW form
\be
ds^2=-dt^2+a(t)^2 \left[\frac{dr^2}{1-k r^2}+r^2\left(d\theta^2+\sin^2 \theta d\phi^2\right)\right],
\ee
where $k$ is the spatial curvature, $a(t)$ is the scale factor, $t$ is the cosmological time and $(r,\theta,\chi)$ are comoving coordinates. It's straightforward to show
that the energy density connected to ${\cal L}_m$ is given by
\begin{eqnarray}
\label{eq3}
\rho_m = \sum_{i}\rho_i+\frac{1}{2}(\dot{\phi}^2+m^2\phi^2)+3\xi\Big[H+\frac{k}{a^2}\Big]\phi^2+6\xi H\phi\dot{\phi},
\end{eqnarray}
with the $\dot{a}\equiv da/dt$ and $H\equiv \dot{a}/a$.

\par
By considering the local Hamiltonian constraint in the Non-Projectable version of Ho\v{r}ava-Lifshitz gravity\cite{sotiriou1}, we obtain the following first integral

\begin{eqnarray}
\label{eq4}
\frac{\dot{a}^2}{2}+\frac{2}{3\lambda-1}\Big\{\frac{\alpha_3 k^3}{6a^4}+\frac{\alpha_2 k^2}{2a^2}+\frac{k}{2}-\frac{\Lambda}{6}a^2
-\frac{4\pi G}{3}\rho_m a^2\Big\}=0.
\end{eqnarray}
where $\alpha_2$ and $\alpha_3$ are constants coupling coefficients. From (\ref{eq4}) we notice that the correction term proportional to $\alpha_2$ behaves just like a radiation component.
\par
Let $T^{\alpha\beta}_{m}$ be the energy momentum tensor connected to the matter
content connected to ${\cal L}_m$.
As in this case the conservation equations $\nabla_{\alpha}T^{\alpha\beta}_{m}=0$ still apply, we obtain
\begin{eqnarray}
\label{eq14.1}
\dot{\rho}_i+3H(\rho_i+p_i)=0 \rightarrow \rho_i=\frac{E_i}{a^{3(1+\omega_i)}}
\end{eqnarray}
for the noninteracting perfect fluids, where $E_i$ are constants of motion. On the other hand, for the nonminimal coupled scalar field we obtain the
following equation of motion

\begin{eqnarray}
\label{eq6}
\ddot{\phi}+3H\dot{\phi}+m^2\phi+6\xi\Big[\frac{\ddot{a}}{a}+H^2+\frac{1}{a^2}\Big]\phi=0.
\end{eqnarray}

\par

If $\omega_i < 1$ for all $i$, the conditions for the bounce are given by
\begin{eqnarray}
\label{eq12}
\lambda>1/3~,~\alpha_3 k^3>0.
\end{eqnarray}
From now on I will restrict myself to the case of closed geometries, that is $k=1$. It is worth to mention that this model is not the only possibility
to generate a bounce in non-relativistic theories. In fact, it has been shown -- for the case closed of FLRW geometries -- that
the Universe can undergo through a bounce as long as the terms which violate relativity lead to a dark radiation component with negative
energy density \cite{saridakisnew}.
%
\par
In order to simplify the numerical analysis I will also fix
$\lambda=1$ and $\alpha_3>0$. In the framework of Ho\v{r}ava-Lifshitz gravity, $\lambda$ must be close to unity
in order to assure that severe Lorentz
violation do not to occur. Thus, the assumption of $\lambda=1$ enables the following model
to be a fair approximation derived from of a healthy non-projectable Ho\v{r}ava-Lifshitz theory.
On the other hand, $\alpha_3>0$ is the necessary and sufficient condition for the bounce.
\par
By choosing the so-called conformal coupling $(\xi=1/6)$, equations (\ref{eq4}) and (\ref{eq6}) may be rewritten using the conformal time $\eta=\int\frac{1}{a}dt$
as
\begin{eqnarray}
\label{eq7}
3{a^{\prime}}^2+V(a)-\frac{1}{2}[{\varphi^{\prime}}^2+(1+m^2a^2)\varphi^2]=8\pi G (E_{rad}-\alpha_4),
\end{eqnarray}
and
\begin{eqnarray}
\label{eq8}
{\varphi^{\prime\prime}}+(1+m^2a^2)\varphi=0,
\end{eqnarray}
where $E_{rad}$ corresponds to a constant connected to $\omega_{rad}=\frac{1}{3}$, the primes denote derivatives with respect to conformal time, $\varphi\equiv a \phi\sqrt{8\pi G} $, $\alpha_4\equiv 3\alpha_2/8\pi G$ and
\begin{eqnarray}
\label{eq8.1}
V(a)=\frac{\alpha_3}{a^2}+3a^2-\Lambda a^4
-8\pi G\sum_{i\neq rad}\frac{E_{i}}{a^{(3\omega_i-1)}}.
\end{eqnarray}

It is worth noting that for $m=0$ the system of equations (\ref{eq7}) and (\ref{eq8}) is separable and, therefore, integrable. That is, in this case, equation (\ref{eq8}) has a first integral ${\cal E}^{0}_{\varphi}=\frac{1}{2}({{\varphi}^{\prime}}^2+\varphi^2)$ which is a constant of motion and, from (\ref{eq7}), we obtain
\begin{eqnarray}
\label{eqII.6}
\eta\equiv\int\sqrt{\frac{3}{8\pi G (E_{rad}-\alpha_4)+{\cal E}^{0}_{\varphi}-V(a)}}da.
\end{eqnarray}
\section{The Structure of the Phase Space}
By considering equations (\ref{eq7}) and (\ref{eq8}), it can be defined the following dynamical system:
\begin{eqnarray}
\label{eq9}
\varphi^\prime=p_{\varphi}~,\\
a^\prime=\frac{p_a}{6}~,\\
p^\prime_\varphi=-(k+m^2a^2)\varphi~,\\
p^\prime_a=-6a+4\Lambda a^3+\frac{2\alpha_3}{a^3}
+{8\pi G}\Big[\sum_i\frac{E_i}{a^{3\omega_i}}(1-3\omega_i)+m^2a\varphi^2\Big].
\end{eqnarray}
Eqs. (\ref{eq9}) and (19) are mere redefinitions. On the other hand, $(a,p_a)$
can be shown to be canonically conjugated by considering the first integral (\ref{eq7}) as a Hamiltonian constraint.
Now we focus on three basic structures that organize the dynamics in the
phase space of the above dynamical system.
\subsection{Invariant Plane}
%
Let us consider the arbitrary dynamical system with $n$ degrees of freedom $(\psi_1,.., \psi_n)\in\mathbb{R}^n$,
whose differential equations are given by
\begin{eqnarray}
\label{mod1.2.1}
\frac{d\psi_i}{d\xi}=F_i(\psi_1,.., \psi_n)~,
\end{eqnarray}
where $i=1,.., n$. Let $\psi_i(\xi)$ be its solution for given initial conditions $\psi_i(\xi_0)\in{\cal M}\subset \mathbb{R}^n$.
Thus, ${\cal M}$ is defined as an invariant manifold if the solution
$\psi_i(\xi)\in {\cal M}$, for each $\psi_i(\xi_0) \in {\cal M}$.
\par
If we fix the initial conditions $p_{\varphi 0}=0=\varphi_0$ we see from equations (\ref{eq9})-(21) that the
dynamics is integrable and does not evolve in the $\varphi$ and ${\varphi}^{\prime}$ directions. That is,
orbits with these initial conditions remain contained in the plane ($a,p_a$) during all the evolution
of the system. Therefore the invariant plane is defined by
\begin{eqnarray}
\label{eqII.11}
{\varphi}=0=p_{\varphi}.
\end{eqnarray}
\par
It is worth noting that the dynamics in this plane is analogous to that of the dynamics in the separable case $m=0$. In fact, in both cases
the dynamics is separable and integrable, and its description is given by similar orbits which differ by a constant ${\cal E}^{0}_{\varphi}$
in the ($a, p_{a}$) sector. As we shall see, in order to furnish an exit to a de Sitter attractor due to parametric resonance I will always
assume initial conditions sufficiently close to the invariant plane.
\subsection{Critical Points}
%
In the phase space $(\psi_1,.., \psi_n)$ of an arbitrary dynamical system like (\ref{mod1.2.1}),
a critical point $\psi^{cr}_{i}$ is defined as a solution of the equations $F_{i}(\psi_1^{cr},.., \psi^{cr}_n)=0$.
That is, it is a stationary solution of (\ref{mod1.2.1}). If one takes the initial condition $\psi_{i}(\xi_0)=\psi^{cr}_{i}$, then $\psi_{i}(\xi)=\psi^{cr}_{i}$ for all $\xi$.
\par
The structure of the phase space of (\ref{eq9})-(21) allows the presence of critical points
$P=(\varphi=0, a=a_{cr}, p_{\varphi}=0, p_a=0)$, where $a_{cr}$ satisfies the relation
\begin{eqnarray}
\label{eq13}
V^{\prime}(a_{cr})\equiv\frac{dV}{da}\Big|_{a=a_{cr}}=0.
\end{eqnarray}
It's easy to see that, by definition, the critical points are contained in the invariant plane.
Furthermore, according to (\ref{eq13}) the critical points are associated to potential extrema.
For specific numerical values of $\Lambda$, $E_{i}$ and $\omega_i$, we may obtain one or many extrema for $V(a)$ (characterized by one or many values of $a_{cr}$). In fact, that is the case for a fixed value of $\Lambda$ and suitable domains of $E_i$.
For $\Lambda=0$, the dynamical system (\ref{eq9})-(21) has only one critical point connected to a global minimum of the potential $V(a)$.
As an exit to the de Sitter attractor can not be performed in this case, I will not consider such configurations.
%
\par
Linearizing the dynamical system (\ref{mod1.2.1}) around the critical points we obtain
\begin{eqnarray}
\label{mod1.2.2}
\frac{d\psi_i}{d\xi}\simeq \sum^{n}_{j=1}dF_{ij}(\psi_j-\psi^{cr}_j),
\end{eqnarray}
where
\begin{eqnarray}
\label{mod1.2.2}
dF_{ij}\equiv \left(
\begin{array}{cccc}
\frac{\partial F_1}{\partial \psi_1} & .. & .. & \frac{\partial F_1}{\partial \psi_n}  \\
:   & .. & .. & : \\
: & .. & .. & :  \\
\frac{\partial F_n}{\partial \psi_1} & .. & .. & \frac{\partial F_n}{\partial \psi_n}  \\
\end{array}
\right)
\end{eqnarray}
Let us then assume that the matrix $dF_{ij}$ has $\mu_l=i\vartheta_l$ pure imaginary eigenvalues (where $l=1,.., 2 p$, $p$ being an integer smaller than $n/2$)
and $\nu_s=\lambda_s$ (where $s=2 p+1,.., n$) real eigenvalues. The respective eigenvectors are given by $\mathbf{x}_l=\mathbf{w}_{~l}+i\mathbf{q}_{~l}$
and $\mathbf{v}_s$. Therefore we define the following
local subspaces (in a small neighbourhood of the critical point) in the phase space
\begin{eqnarray}
\label{mod1.2.3}
W^s=Span\{\mathbf{v}_s|\lambda_l<0\}, \\
W^u=Span\{\mathbf{v}_{s}|\lambda_l>0\}, \\
W^c=Span\{\mathbf{w}_{~l},\mathbf{q}_{~l}|\lambda_l=0\}.
\end{eqnarray}
The superscript $s$, $u$ and $c$ denote {\it stable}, {\it unstable} and {\it center} manifolds respectively.
If $dF_{ij}$ has only pure imaginary eigenvalues, the critical point is called a {\it center}.
If $dF_{ij}$ has $n-2$ pure imaginary eigenvalues and two real eigenvalues (one positive and one negative),
the critical point is called a {\it saddle-center}.

\par
In the case of the dynamical system (\ref{eq9})-(21), we obtain
%
\begin{eqnarray}
\label{eqII.16}
\left(
\begin{array}{c}
\varphi^\prime  \\
p^\prime_\varphi   \\
a^\prime   \\
p^\prime_a   \\
\end{array}
\right)=\left(
\begin{array}{cccc}
0 & 1 & 0 & 0  \\
-(1+m^2 a^2_{cr})   & 0 & 0 & 0 \\
0 & 0 & 0 & \frac{1}{6}  \\
0 & 0 & -\frac{\partial^2 V(a)}{\partial a^2}|_{a=a_{cr}} & 0  \\
\end{array}
\right)
\left(
\begin{array}{c}
\varphi  \\
p_\varphi   \\
a-a_{cr}   \\
p_a   \\
\end{array}
\right).
\end{eqnarray}
It can be shown that the above matrix has the following eigenvalues
\begin{eqnarray}
\label{eq14}
\mu_{1,2}=\pm i \sqrt{1+m^2a^2_{crit}}~,~\mu_{3,4}=\pm\sqrt{-\frac{V^{\prime\prime}(a_{crit})}{6}}.
\end{eqnarray}
By considering the plane $(\varphi, p_{\varphi})$, the corresponding eigenvectors of $\mu_1$ and $\mu_2$ engender the local topology
of $S^1$ around the origin $(\varphi=0, p_{\varphi}=0)$.
Thus, the local topology of the critical points is determined by the second derivative of the potential $V(a)$.
\par
When $\Big(\frac{\partial^2 V(a)}{\partial a^2}\Big)\Big|_{a=a_{cr}}>0$
we obtain a local minimum for the potential $V(a)$ and one more pair of pure imaginary eigenvalues.
In this case, by considering the plane $(a, p_{a})$, the corresponding eigenvectors of $\mu_3$ and $\mu_4$ engender the local topology
of $S^1$ around the point $(a=a_{cr}, p_{a}=0)$.
As a consequence, the topology around the critical points with $\Big(\frac{\partial^2 V(a)}{\partial a^2}\Big)\Big|_{a=a_{cr}}>0$
is given by $S^1 \times S^1$ (or 2-torus). We denote by $P_0$ a critical point with this property. It is called a {\it center} by definition.
\par
When $\Big(\frac{\partial^2 V(a)}{\partial a^2}\Big)\Big|_{a=a_{cr}}<0$
we obtain a local maximum for the potential $V(a)$ and two real eigenvalues (one positive and one negative).
In this case, by considering the plane $(a, p_{a})$, the corresponding eigenvectors of $\mu_3$ and $\mu_4$ engender the local topology
of a {\it saddle} around the point $(a=a_{cr}, p_{a}=0)$.
As a consequence, the topology around the critical points with $\Big(\frac{\partial^2 V(a)}{\partial a^2}\Big)\Big|_{a=a_{cr}}<0$
is given by $R\times S^1$. We denote by $P_1$ a critical point with this property. It is called a {\it saddle-center} by definition.
\par
The expansion of the first integral (\ref{eq7}) around the critical points reads

\begin{eqnarray}
\label{eqII.18}
\nonumber
H\equiv \frac{1}{12}p_a^{2}+\frac{1}{2} V^{\prime \prime}(a_{crit})~ (a-a_{crit})^2
-\frac{1}{2}[p_{\varphi}^{2}+(1+m^2 a_{\rm crit}^2)\varphi^2]+E_{ crit}\\
-8\pi G(E_{rad}-\alpha_4)+
{\cal{O}}(3)=0,
\end{eqnarray}
where ${\cal{O}}(3)$ denote terms of higher order in the expansion and $E_{{crit}}\equiv V(a_{crit})$ is the energy of the respective critical point.
In a small neighborhood of the critical points these higher order terms can be neglected and the dynamics is nearly separable in the sectors ($a,p_a$)
and ($\varphi, p_{\varphi}$) with constants of motion given by
\begin{eqnarray}
\label{eqII.19}
E_{(a)}= \frac{1}{12}p_a^{2}+\frac{1}{2} V^{\prime \prime}(a_{\rm crit})~ (a-a_{\rm crit})^2~,\\
E_{(\varphi)}= \frac{1}{2}\Big[p_{\varphi}^{2}+(1+m^2 a_{\rm crit}^2)\varphi^2 \Big],
\end{eqnarray}
with $E_{(a)}-E_{(\varphi)}+ E_{crit}-8\pi G(E_{rad}-\alpha_4) \sim 0$ and $|E_{crit}-8\pi G(E_{rad}-\alpha_4)|$ sufficiently small.
While in the sector $(\varphi,p_{\varphi})$ we have a rotational motion with energy
$E_{(\varphi)}$ around the critical points, in the sector $(a, p_{a})$ we have two possibilities.
If $V^{\prime \prime}(a_{cr})>0$ we have a rotational motion with energy $E_{(a)}$ in a small neighborhood of the critical point
$P_{0}$. Otherwise ($V^{\prime \prime}(a_{cr})<0$), we obtain a hyperbolic motion around the critical point $P_{1}$.
The critical point $P_1$ defines an universe analogous to that of the unstable Einstein static universe \cite{wald}.
On the other hand, the configuration of stable Einstein static universe, corresponding to the critical point $P_0$, does not possess any
classical analogue.
\subsection{Separatrices}
%
According to the definition above, from the {\it saddle-center} emerges a special structure consisting in two local subspaces. While one of them is generated by the unstable manifold $W^u$,
the other is generated by the stable manifold $W^s$. Being transversal to each other they define the separatrices of the {\it saddle-center} critical point.
\par
Let us then assume that one of the initial conditions is given by a point in the phase space which lies on $W^u$ in a neighbourhood of the {\it saddle-center} critical point. Although this manifold is locally unstable,
in general this does not mean that the final stage of the dynamics differs from the {\it saddle-center} critical point. In fact, sometimes the nonlinearities of the dynamics may induce an orbit to join the {\it saddle-center} point to itself. In this case, the final stage of the dynamics is the the very same {\it saddle-center} critical point. Orbits with such a property are called homoclinic orbits.
\par
From the saddle-center critical point $P_1$ (when
present) emerges a structure of separatrices S contained
in the invariant plane. One of them tends to a
de Sitter attractor at infinity, defining an escape of orbits
to an accelerated phase regime. In fact, a straightforward analysis
of the infinity of the phase space shows the presence of a
pair of critical points in this region, one acting as an
attractor (stable de Sitter configuration) and the other as
a repeller (unstable de Sitter configuration). The scale
factor approaches the de Sitter attractor as $a(\eta)\sim (C_0-\eta)^{-1}$ for $\eta \rightarrow C_0$,
or $a(t) \sim {\rm exp} \Big(t \sqrt{\Lambda/3} \Big)$.
\section{A Simple Model}
Let us now consider that the noninteracting perfect fluids are given by dust and radiation.
In this case it can be shown that the potential
\begin{eqnarray}
\label{eq15.0}
V(a)=3a^2-\Lambda a^4
-8\pi G E_{dust} a + \frac{\alpha_3}{a^2}
\end{eqnarray}
will have two extrema (one local minimum
and one local maximum) as long as the following conditions hold:
\begin{eqnarray}
\label{eq15}
27(8\pi G E_{dust})^2>\frac{256}{\Lambda},
\end{eqnarray}
and
\begin{eqnarray}
\label{eq16}
6\gamma-4\gamma^3\Lambda-8\pi G E_{dust}>\frac{2\alpha_3}{\gamma^3},
\end{eqnarray}
where
\begin{eqnarray}
\label{eq17}
\nonumber
\gamma=\frac{\rho}{12\Lambda}+\frac{4}{\rho}~,\\
\nonumber
\rho^3=\Big[-108(8\pi G E_{dust})+12\sqrt{3}\sqrt{27(8\pi G E_{dust})^2-\frac{256}{\Lambda}}\Big]\Lambda^2.
\end{eqnarray}
It can be numerically shown (cf. Fig. 1) that the increase $E_{dust}$ has the effect of spoiling the potential configuration
with two extrema.

If $m=0$ the dynamics is integrable and thus separable. In this case the first integral (\ref{eq7}) reads
\begin{eqnarray}
\label{eq15}
\frac{p^2_a}{12}+{V}(a)=8\pi G (E_{rad}-\alpha_4)+{\cal E}^0_{\varphi}\equiv E^0_{a},
\end{eqnarray}
and the scalar field behaves just like a radiation component in the dynamics of the scale factor.
\par
In Fig. 2 I exhibit the phase portrait in the invariant plane $\varphi=0=p_{\varphi}$.
With suitable values for the parameters, the two extrema of $V(a)$ are connected to $P_0$ and $P_1$.
This model furnishes us with perpetually bouncing universes (periodic orbits in region I).
The two separatrices $S_1$ and $S_2$ that emerge from $P_1$ coalesce generating a boundary for region I.
This boundary defines an homoclinic orbit by definition.
Orbits in region II and III correspond to universes with one bounce only.
\begin{figure}
\begin{center}
\includegraphics[height=7cm,width=11cm]{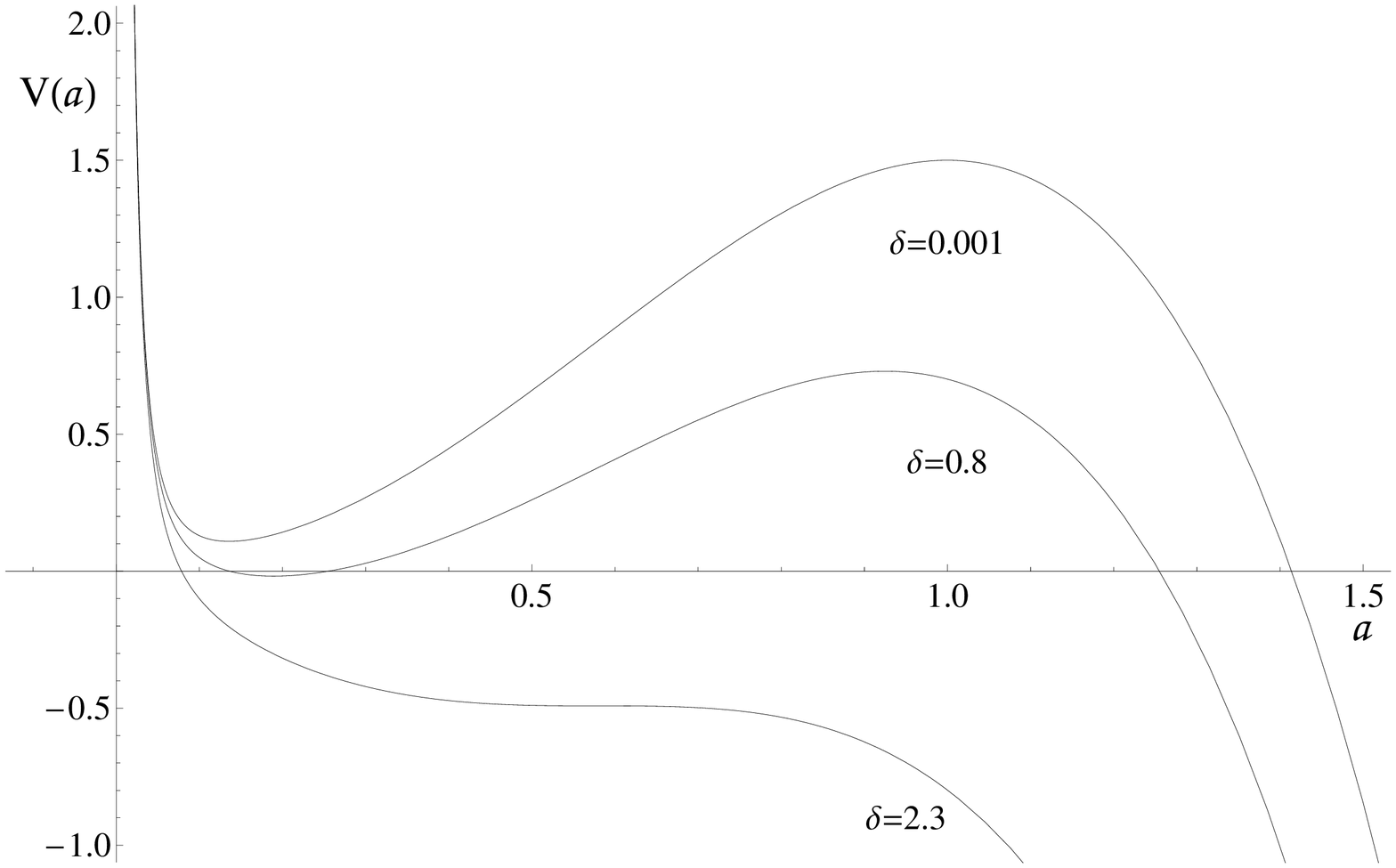}
\caption{The potential ${V}(a)$ for several values $\delta \equiv E_{dust}$.
For higher values of $\delta$ the minimum of the potential is no longer present. For $\delta \simeq 2.3$
the two extrema of the potential (connected to the stable and unstable Einstein static universes) vanish.
Here we fixed $\Lambda=1.5$, $\alpha_3=10^{-3}$ and $8\pi G = 1$.}
\vspace{0.5cm}
\includegraphics[height=7cm,width=11cm]{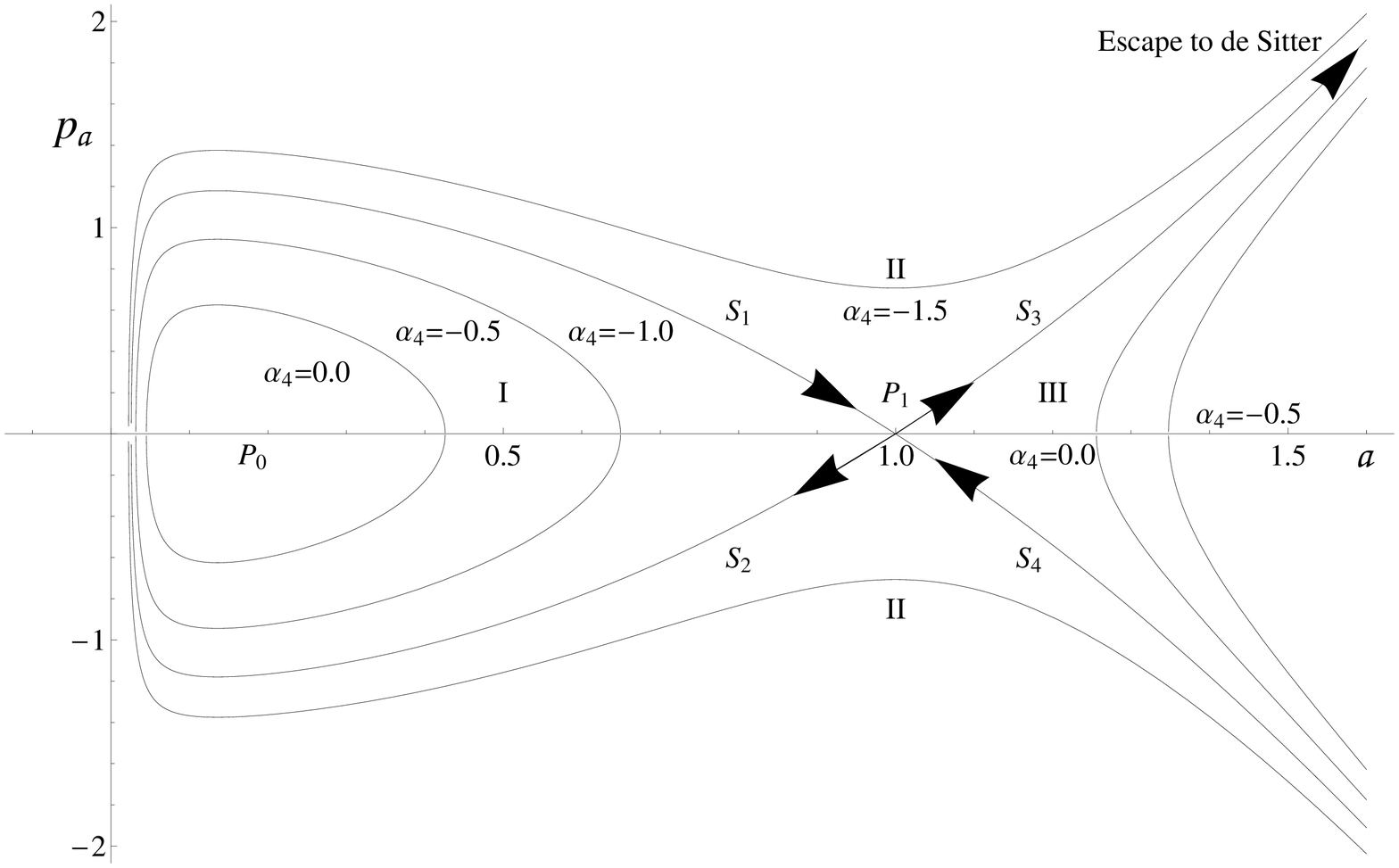}
\caption{The phase portrait of invariant plane dynamics with the critical points $P_0$ (center)
and $P_1$ (saddle-center) corresponding to stable and unstable Einstein universes.
The periodic orbits of region I describe perpetually bouncing universes.
Orbits in Region II and III are solutions of one-bounce universes. A separatrix $S_3$ emerges
from $P_1$ defining an escape to the de Sitter attractor. Here we fixed $\Lambda=1.5$, $\alpha_3=10^{-3}=E_{dust}$, $E_{rad}=0.5$ and $8\pi G = 1$.}
\label{fig1}
\end{center}
\end{figure}

\par

Now I focus on some structural differences between the integrable dynamics in the invariant plane
and the integrable dynamics given by $m=0$. If $\varphi(0)$ and/or  $p_{\varphi}(0)$
do not vanish (in the integrable case $m=0$), the phase portrait in the plane ($a,p_a$) is analogous to that of the invariant plane (cf. Fig. 2).
Let $X_0$ and $X_1$ be the analogous that of $P_0$ and $P_1$ (connected to the potential extrema).
In this case, $X_0$ and $X_1$ would be no longer critical points. Instead, they define stable and unstable periodic orbits respectively.
Due to the system's periodicity in a neighborhood of $P_0$, orbits are confined on two dimensional surfaces
in the phase space which topologically coincide with $2$-tori. Therefore, the integrable dynamics ($m=0$) is not constrained in the invariant plane but,
in a neighborhood of $P_0$, evolves on $2$-tori which are the direct product of closed curves (analogous to that of in region I) with periodic orbits in the sector
($\varphi, p_{\varphi}$).

\par

Another topological feature to be considered -- in the integrable case $m=0$ -- is the dynamics of orbits in a neighborhood of $P_1$.
Given the initial conditions $p_{a}(0)=0$ and $a_0=a|_{X_{1}}$, the motion of the system corresponds to an unstable periodic orbit
which generates a circle in the sector ($\varphi, p_{\varphi}$). Let $\Gamma$ denotes one of these orbits.
Therefore, the direct product of periodic orbits
in the sector ($\varphi, p_{\varphi}$) with the stable and unstable manifolds (analogous to that of $S_1$ and $S_2$) generates semi-cylinders which coalesce in the very same
orbit $\Gamma$. When we consider a nonlocal description of the topology, the nonlinearities induce such cylinders to close onto themselves.
As the projection of such cylinders in the sector ($a,p_a$) defines an homoclinic orbit, we call these structures as homoclinic cylinders.
On the other hand,
the direct product of periodic orbits in the sector $(\varphi, p_{\varphi})$ with the analogous to that of separatrices $S_3$ and $S_4$
generates semi-infinite cylinders which coalesce in the orbit $\Gamma$. A summary of structural differences between the dynamics in the invariant plane
and the integrable case $m=0$ is given in the table below.
\newline
\begin{center}
\begin{tabular}{|c|c|}
\hline
Dynamics in the Invariant Plane: & Integrable Dynamics $m=0$: \\\hline
No motion in the sector ($\varphi, p_{\varphi}$)& Separable motion in the sectors ($a, p_{a}$) and ($\varphi, p_{\varphi}$)\\\hline
Critical points $P_{0}$ e $P_{1}$ & Stable and unstable periodic orbits\\\hline
Periodic orbits (Region I) & Integrable Tori \\\hline
Separatrices between regions I e II & Homoclinic cylinders \\\hline
Separatrices between regions II e III & Semi-infinity integrable cylinders \\\hline
\end{tabular}
\end{center}
\vspace{1cm}
\section{Nonlinear Resonance of KAM Tori}
\par
According to Liouville-Arnold \cite{arnold} theorem, if the motion of a hamiltonian system with $n$ degrees of freedom
is integrable and bounded, then the orbits of such a system are confined on $n$-dimensional hyper-surfaces in the phase
space which topologically coincide with $n$-tori. That is exactly what occurs with our system when we consider the
integrable case $m=0$ in a neighborhood of $P_0$.
\par
In order to give a more precise analysis, let us consider the surfaces
with energy $E^{0}_{a}=8\pi G (E_{rad}-\alpha_4)+ {\cal E}_{\varphi}^{0} < E_{crit}(P_1)\equiv V(P_1)$.
This region of the phase space is foliated by $2$-tori ${\mathcal{S}}^1 \times {\mathcal{S}}^1$ which are the topological
product of periodic orbits of the separable sectors ($\varphi,p_{\varphi}$) and ($a,p_a$). ${\cal E}_{\varphi}^{0}$ and ${E}_{a}^{0}$
are conserved quantities for those orbits. The frequency $\nu_a$ of the periodic orbit in the sector ($a,p_a$) is given by the
integral
\begin{eqnarray}
\label{eqII.26}
\frac{1}{\nu_a}=\sqrt{\frac{12}{\Lambda}}\int_{\beta_1}^{\beta_2}\frac{a~da}{\sqrt{\prod_{i=1}^{6}(a-\beta_i)}},~~
\end{eqnarray}
where $\beta_i$ ($i=1...6$) are the real roots of ${V}(a)=8\pi G (E_{rad}-\alpha_4)+{\cal E}_{\varphi}^{0}$.
Here I denote $\beta_i$ ($i=1...3$) as the three positive real roots with $\beta_1<\beta_2<\beta_3$. On the other hand,
the periodic orbits in the sector ($\varphi,p_{\varphi}$), parameterized by ${\cal E}_{\varphi}^{0}$, have the frequency
$\nu_{\varphi}=1/2\pi$.

The importance of $n$-tori in integrable hamiltonian systems comes from the fact that these surfaces trap the dynamics
in a finite region of the phase space. In our case, such tori in a neighborhood of $P_0$ avoid an exit to the de Sitter attractor.
However, a relevant question which arises is whether such tori ``survive'' when one introduces a small perturbation connected
to the mass $m$ of the scalar field. Assuming initial conditions ($\varphi_0,p_{\varphi0}$) sufficiently close to the invariant plane,
equation (\ref{eq8}) may be rewritten as

\begin{eqnarray}
\label{eqII.27}
{\varphi}^{\prime\prime}+(1+m^2{a_0^2(\tau)})~ \varphi=0,
\end{eqnarray}
where $a_0(\eta)$ is the background solution for the scale factor of the integrable dynamics (\ref{eqII.6}).
Equation (\ref{eqII.27}) is a Lam\'e equation. Defining ${\tilde{\nu}_{\varphi}}$ as the frequency in the
($\varphi, p_{\varphi}$) given by (\ref{eqII.27}), the resonance phenomena \cite{arnold} will occur when the ratio

\begin{eqnarray}
\label{eqII.28}
R\simeq\frac{\nu_a}{{\tilde{\nu}_{\varphi}}}
\end{eqnarray}
is a rational number. Expanding $a_0(\eta)$ in the Lam\'e equation, one can show that

\begin{eqnarray}
\label{eqII.29}
{\tilde{\nu}}_{\varphi}\simeq\frac{1}{2\pi}\sqrt{1+\frac{(0.9~m)^2}{2}~(\beta_{1}^{2}+\beta_{2}^{2})}.
\end{eqnarray}
\par
However, as the system evolves the amplitude of the scalar field may grow so that the solution
of the integrable case $a_{0}(\eta)$ is no longer a good approximation to be introduced in (\ref{eq8}). As we shall see in the next section
this process may lead the dynamics into a more unstable behavior, with the amplification
of the resonance mechanism and the break of the KAM tori \cite{berry}.
To analytically show this behavior, let us now consider the following approximation of constraint (\ref{eq7})
\begin{eqnarray}
\label{eqII.30}
{\mathcal{H}} \equiv {E}_{a}^{0}-{\cal E}_{\varphi}^{0}-\frac{1}{2}{m^2 a_{0}^2(\eta) \varphi^2(\eta)}\simeq 8\pi G(E_{rad}-\alpha_4),
\end{eqnarray}
where $\varphi(\eta)$ is an approximate solution of the Lam\'e
equation. Now I introduce the action-angle variables \cite{arnold} ($\Theta_{\varphi}, {\cal{J}}_{\varphi}, \Theta_a,
{\cal{J}}_a$). The angle variables are defined by ($\Theta_{\varphi}={\tilde{\nu}}_{\varphi}
 \eta,~\Theta_a=\nu_a \eta,$) in such a way that they span the interval $[0,1]$ during a complete period of the system.
Taking into account that the function $a_0(\eta)$ is periodic with period $T_a=\nu_a^{-1}$,
the expansion of the non-integrable term of (\ref{eqII.30}) is given by \cite{abramo}
\begin{eqnarray}
\label{eqII.31}
-\frac{1}{2}{m^2 a_{0}^2(\eta) \varphi^2(\eta)}=-\frac{1}{2}~m^2{\cal{J}}_{a}^{(0)}{\cal{J}}_{\varphi}^{(0)}\sum_{n} \Big(A_n \cos 2 n \pi \Theta_{a}\Big) \cos 4 \pi \Theta_{\varphi},~
\end{eqnarray}
where $A_n$ are constant coefficients.
The superior index in ${\cal{J}}_{a}$ and $\cal{J}_{\varphi}$ denotes that these are the action variables
for the integrable case. The Hamilton equation for ${\cal{J}}_{a}$, derived from (\ref{eqII.30}), can then be integrated
furnishing us in first approximation with
\begin{eqnarray}
\label{eqII.32}
\nonumber
{\cal{J}}_{a} \sim \frac{1}{2}~m^2{\cal{J}}_{a}^{(0)}{\cal{J}}_{\varphi}^{(0)}\sum_{n} \frac{A_n}{2 \pi n {\tilde{\nu}}_{\varphi}}\Big[\frac{\cos (2 \pi n
\Theta_{a}-4 \pi \Theta_{\varphi})}{R-2/n}\\
~~~~~~~~~~~~~~~~~~~~~~~~~~~~~~~~~~~~~~~~~~~~~~~~~~~~~~~~+\frac{\cos (2 \pi n \Theta_{a}+4 \pi \Theta_{\varphi})}{R+2/n} \Big].~
\end{eqnarray}
From (\ref{eqII.32}) we see that the dominant resonance terms are those for which $R \simeq 2/n$.
It can be shown that the condition
$n \geq 2$ must hold in order to obtain a real positive numerical value for the mass $m$. Therefore,
\begin{eqnarray}
\label{eqII.33}
R\simeq\frac{\nu_a}{{\tilde{\nu}_{\varphi}}}\simeq\frac{2}{n},~~~(n \geq 2)
\end{eqnarray}
is a good approximation in order to determine the resonances of the dynamical system (\ref{eq9})-(21) in the presence of dust and radiation. When such resonances occur one can eventually obtain a loss
of stability with the break of the KAM tori \cite{berry}, allowing the dynamics to an exit to the de Sitter attractor.
\par
Let us now consider a Poincar\'e map in the variables ($\varphi, p_{\varphi}$) with section $p_{a}=0$. As a convention I will assume that this map is
unidirectional. That is, a chosen orbit of the system crosses the plane of section $p_a=0$ only once after a period of time $T_{a}$. If
${\cal E}_{\varphi}^{0} = 0$, the periodic orbits in the sector ($a, p_a$) correspond to the point ($\varphi=0, p_{\varphi}=0$).
For ${\cal E}_{\varphi}^{0} \neq 0$ and $m=0$, the tori are characterized by closed curves around the origin of this map. For a small value of $m$ this configuration is maintained around the origin ($\varphi=0, p_{\varphi}=0$). In fact, according to the KAM theorem \cite{kam},
if $\nu_{a}$ and $\nu_{\varphi}$ are sufficiently irrationals (that is, satisfy the diophantine condition) then
the solutions of the perturbed system are quasi-periodic for a sufficiently small value of $m$.
\par
In order to compare the approximation (\ref{eqII.33}) with the exact dynamics I will perform a numerical analysis of the evolution of the system
in the Poincar\'e map with section $p_{a}=0$. For computational simplicity I will fix $\Lambda=1.5$, $E_{rad}=0.5$, $\alpha_3=10^{-3}=E_{dust}$ and $8\pi G =1$. In this way one can define the parametric space ($\alpha_4, m$) where the resonances may occur.
For several initial conditions around ($p_{\varphi_0}=0, \varphi_0=0$) I construct the Poincar\'e map with $m=9.6$ (Fig. 3) and $m=11.5$ (Fig. 4).
According to approximation (\ref{eqII.33}) the first map shows the resonant behavior of the system for $n=3$. The second map shows the pattern
of parametric stability in a region between the resonances $n=3$ and $n=4$.
\begin{figure}
\begin{center}
\includegraphics[width=10cm,height=6.5cm]{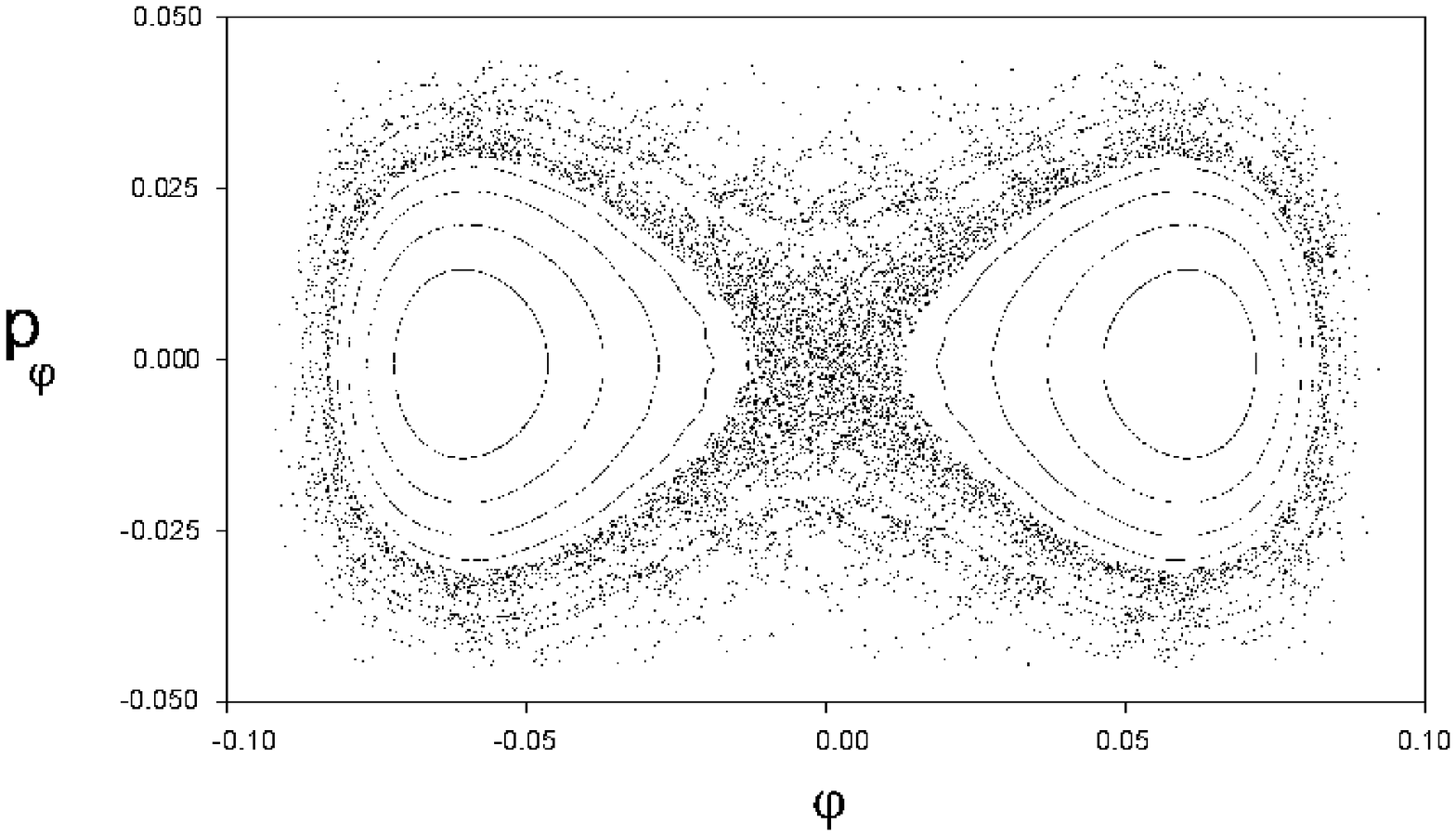}
\caption{Poincar\'e map with section $p_a=0$ for several initial conditions with $m=9.6$ in the domain
of parametric resonance $n=3$. Here we see that the resonance of the exact dynamics is connected to the bifurcation of the stable periodic orbit at the origin
($\varphi=0,p_{\varphi}=0$). That is, when the resonance occurs this stable periodic orbit turns into a unstable periodic orbit. In this case no KAM tori are present around the origin of the map so orbits with initial conditions in a neighborhood of the origin may escape to the de Sitter attractor. The presence of such bifurcation is a crucial feature in order to allow the dynamics to an exit to the de Sitter attractor.}
%
%
\vspace{1cm}
\includegraphics[width=10cm,height=6.5cm]{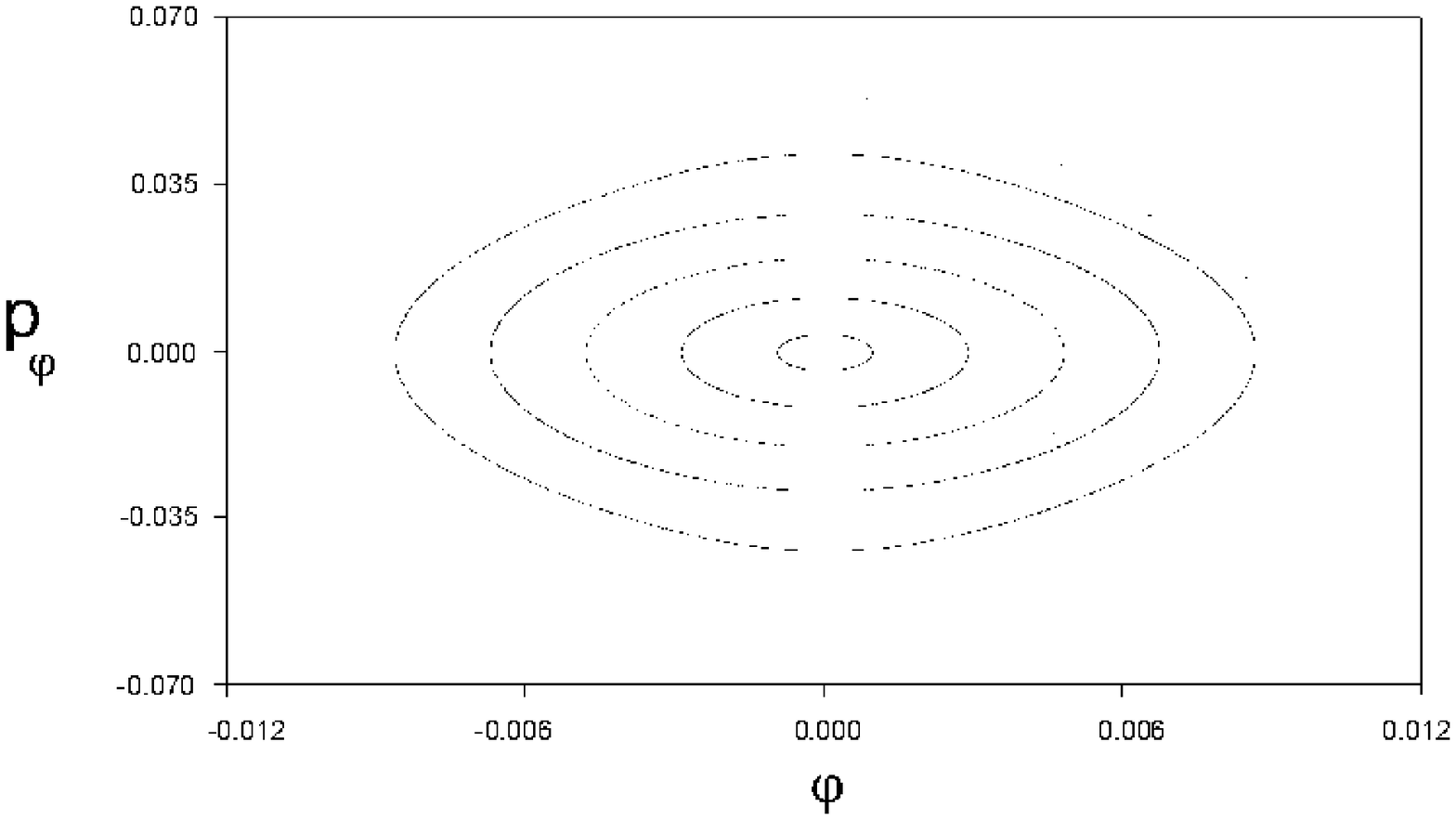}
\caption{Poincar\'e map of section $p_a=0$ for several initial conditions with $m=11.5$ in the domain of parametric stability
between the resonances $n=3$ and $n=4$. The topology of the system around the origin (which defines the stable periodic orbit) corresponds to a $2$-torus. In this case, the remaining KAM tori from the integrable case trap the orbits with initial conditions in a neighborhood of the origin avoiding an escape to the de Sitter attractor.
Therefore the region of parametric stability of the system does not favor late-time accelerating scenarios.}
\end{center}
\end{figure}
The structure of the stochastic sea \cite{chernikov} in Fig. 3 shows that initial conditions
near the invariant plane can generate orbits with a long time of diffusion before escaping to the de Sitter attractor.
\par
In Fig. 5, I numerically construct the resonance chart using the exact dynamics.
Taking the initial conditions $p_{a_{0}}=0=p_{\varphi_{0}}$ and $\varphi_{0}=10^{-3}$, the value
of $a_{0}$ is obtained by substituting the numerical values of $\alpha_4$ and $m$ in the constraint (\ref{eq7}).
For a suitable value of $\alpha_4$, the approximate expression (\ref{eqII.33}) is an accurate guide in order
to localize the respective values of $m$ in which the resonances in the parametric space ($\alpha_4, m$) occur.
The dashed curves (cf. Fig. 5) in the parametric space ($\alpha_4, m$)
are constructed using approximation (\ref{eqII.33}) and they allow us to localize a given domain of resonance.
As I previously pointed out, the dominant resonances of the system are connected to the bifurcation of stable periodic orbits at the
origin. Although approximation (\ref{eqII.33}) allow us identify the curves in the parametric space ($\alpha_4, m$) where the resonances occur, the effect of the exact dynamics
tends to stretch these domains. In fact, as one may numerically verify, for a fixed value of $\alpha_4$ there is a continuum domain of values of $m$ (where the bifurcation of stable periodic orbits at the origin occurs) for each resonance. The windows of exact resonance are shown as hatched regions in Fig. 5.
\begin{figure}
\vspace{0.1cm}
\begin{center}
\hspace{0.0cm}
\vspace{-0.3cm}
\includegraphics*[height=8cm,width=13cm]{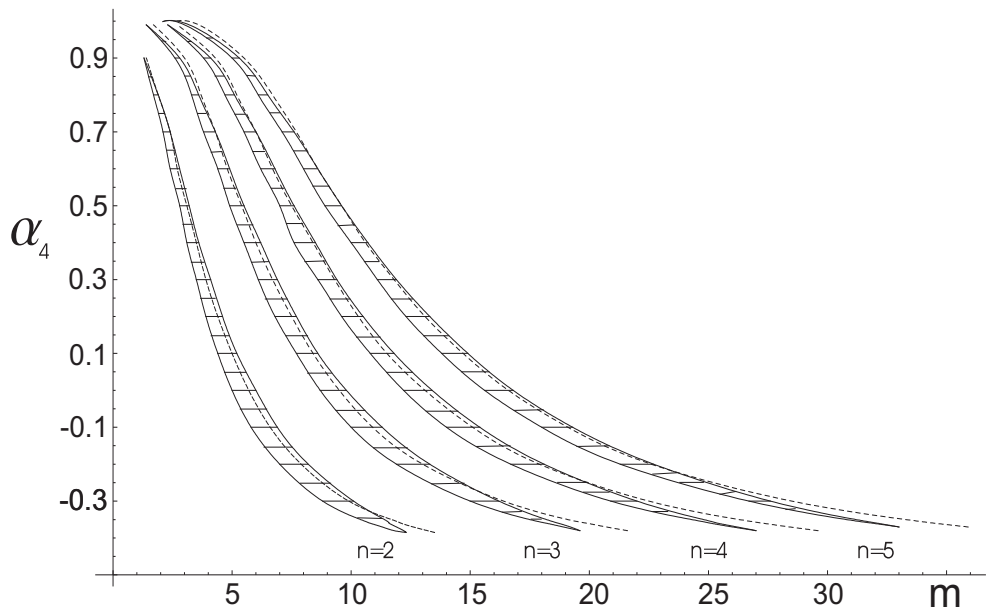}
\caption{Resonance chart in the plane ($\alpha_4,m$). The dashed lines are solutions of the conditions of resonance (\ref{eqII.33}). The hatched areas
correspond to resonance domains of the exact dynamics. These domains correspond to regions where the bifurcation of periodic orbits at the origin (in the Poincar\'e map of section $p_a=0$) occurs. The white remaining region corresponds to the domain of parametric stability where the dynamics is trapped by the KAM tori. For computational simplicity I am still fixing
the parameters $\Lambda=1.5$, $E_{rad}=0.5$, $\alpha_3=10^{-3}=E_{dust}$ and $8\pi G =1$.}
\label{fig3}
\end{center}
\end{figure}
\section{{The Resonance Pattern}}
The resonance regions in the parametric space $(\alpha_4, m)$ possess a substructure
which I now examine. In order to simplify this analysis I will restrict myself to the resonance domain $n=3$ (cf. Fig. 5)
with $\alpha_4=0$. For these fixed values the resonance occurs in the interval $\Delta m \cong [8.7,9.8]$.
In this interval one can notice three distinct regions.
\par
(i) For $m < 9.3$ the dynamics is highly unstable and the resonances provide a rapid escape to the de Sitter attractor.
In Fig. 6 the behavior of $a$ and $\varphi$ (with respect to the conformal time) is shown
by taking $m=8.8$. In this figure one can observe that a rapid escape to the de Sitter attractor occurs when $\eta \simeq 70$
so there is no enough recurrence in order to construct a Poincar\'e map.

\begin{figure}
\vspace{0.2cm}
\begin{center}
\hspace{0.0cm}
\vspace{0.0cm}
\includegraphics*[height=5cm,width=12cm]{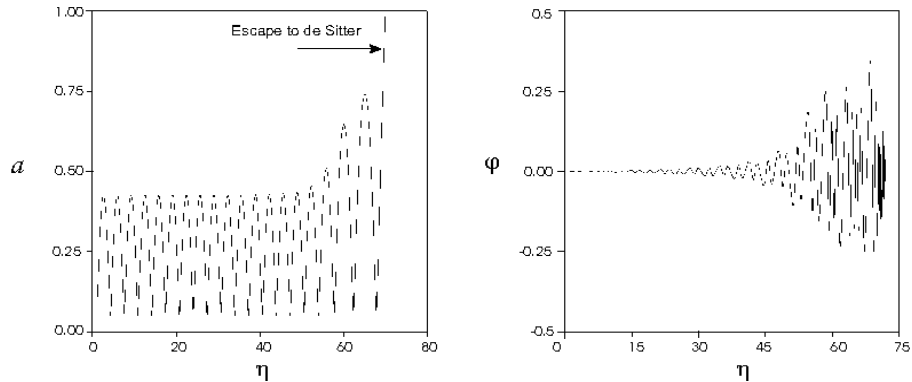}
\caption{The evolution of the scale factor and the scalar field for $m=8.8$.
This behavior characterizes a region of disruptive resonance close to the left edge of resonance $n=3$ of Fig. 5.}
\label{fig2.6}
\end{center}
\end{figure}
\par
(ii) When $9.7 < m < 9.8$ the motion of orbits is resonant and chaotic, although stable.
In Fig. 7 the behavior of an orbit in this region is exhibited for $m=9.8$. This is the pattern close
to the right edge of resonance $n=3$. Due to its stability these orbits are not interesting from the
late-time accelerating point of view.
It is worth mentioning that this behaviour is in agreement to
the analytical description regarding cyclic universes given in \cite{saridakisnew}.
%
\begin{figure}
\vspace{0.1cm}
\begin{center}
\hspace{0.0cm}
\vspace{0.0cm}
\includegraphics*[height=5cm,width=12cm]{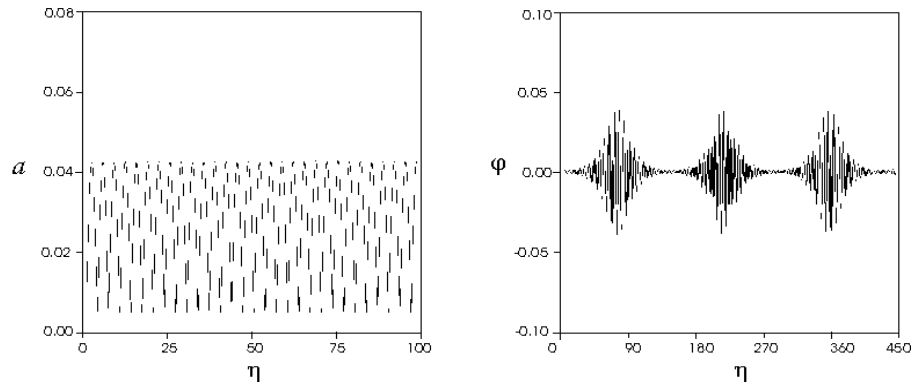}
\caption{The evolution of $a$ and $\varphi$ for $m=9.8$
(corresponding to the right edge of resonance zone $n=3$ of Fig. 5). The motion in this region is not
interesting from the late-time accelerating point of view.}
\label{fig2.7}
\end{center}
\end{figure}
%
%
\par
(iii) A region of transition occurs when $9.3 < m \leq 9.7$. In this case, orbits go through a long time
of diffusion before escaping to the de Sitter attractor.
In Fig. 8 a Poincar\'e map (with $p_a=0$) of an orbit with $m=9.7$ is shown. This map illustrates
what happens in the above interval. This is an example of how orbits can go through a long time of diffusion before
escaping to the de Sitter attractor.
\begin{figure}
\vspace{0.2cm}
\begin{center}
\hspace{0.0cm}
\vspace{-0.3cm}
\includegraphics*[height=6cm,width=9.5cm]{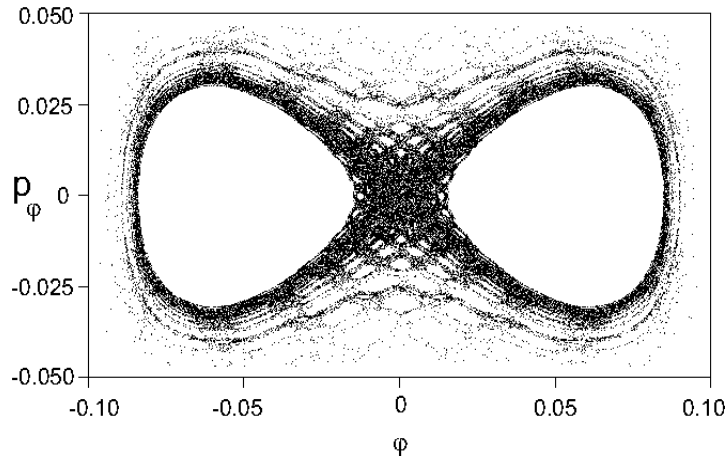}
\caption{Poincar\'e map (with section $p_a=0$) of an orbit with $m=9.7$.
This is the behavior in the region of transition of resonance $n=3$ of Fig 5. In this case one can observe that the orbit go through a long time
of diffusion before escaping (when $\eta \simeq 10^6$) to the de Sitter attractor. This map exhibits a dark region connected to the structure
of random motion of the orbit in a stochastic sea around the KAM islands.}
\label{fig2.8}
\end{center}
\end{figure}
\par
The above substructure is a pattern which is maintained for every value of $\alpha_4$ in the resonance zone $n=3$. Furthermore, it can be shown that this pattern is qualitatively equivalent for every value of $n$.
Throughout this analysis one notices that nonsingular perpetually bouncing models from Ho\v{r}ava-Lifshitz possess a restrict domain
in the parametric space where late-time acceleration (connected to the de Sitter attractor) may be realized. For typical variations of the parameters $E_{dust}$ and/or $\alpha_3$ the domains of the parametric space ($\alpha_4, m$) -- where the system is resonant -- can be stretched or shrunken. Nevertheless
the pattern in resonance windows and its substructure are maintained as one may numerically verify. In this sense the
pattern is said to be structurally stable.

\section{Conclusions}

In this paper I examine the effect of parametric resonance in bouncing cosmologies originating from Ho\v{r}ava-Lifshitz gravity.
In this context, terms arising from foliation preserving diffeomorphism invariance -- which breaks $4$-D covariance --
implement nonsingular bounces in the early evolution of the universe.
The matter content of the model is given by perfect fluids, namely dust and radiation. Furthermore I also assume a nonvanishing cosmological constant
(connected to a de Sitter attractor in the phase space which provides
late-time acceleration) and a massive conformally coupled scalar field.

By considering the case of closed
geometries I obtain a potential well with a local minimum and a local maximum (cf. Fig. 1) respectively connected to the critical points
$P_0$ (center) and $P_1$ (saddle-center) in the phase space. Assuming a conformally coupled scalar field, the oscillatory behavior of the dynamical system
around $P_0$ might become metastable when the system is driven into a resonance
window of the parameter space -- labeled by an integer $n\geq2$. In this case I determine the physical domain of the parameters (cf. Fig. 5)
in which the breakup of KAM tori may occur, leading the Universe to a late-time acceleration regime.

It is worth mentioning that, as examined in \cite{maier}, a chaotic exit to accelerated expansion can be also realized
-- in the dynamical system (\ref{eq9})-(21) -- if one assumes
initial condition sets taken in a small neighborhood of the stable
separatix $S_1$. These sets possesses fractal basin
boundaries connected to a code recollapse/escape leading to a chaotic exit to an accelerated regime.

Although the cosmological constant poses a crucial problem to quantum
field theory on how to match its observed value
with vacuum energy calculations, the cosmological constant is
by far the simplest explanation for the present acceleration
of the Universe. Indeed, the $\Lambda$CDM standard model assumes that there exists a cosmological
constant which becomes dynamically important when the typical scale of the
Universe has the size of the present Hubble radius.
In this sense, the model of this paper does not exhibit an alternative explanation for late-time acceleration.
Instead, the core of this paper is to examine the dynamics in the phase space of the above model, showing how to provide
an alternative exit to late-time acceleration.

\section{Acknowledgements}

I acknowledge financial support
of CNPq/MCTI-Brazil, through a Post-Doctoral
Grant No. 201907/2011-9. I would like to thank
David Wands for his useful comments
and suggestions. I also would like to acknowledge the Institute of Cosmology and Gravitation for their hospitality. Figures were generated
using the Wolfram Mathematica $7$ and DYNAMICS SOLVER packet \cite{ds}.

\section*{References}

\end{document}